\documentstyle[aps,preprint]{revtex}
\tightenlines

\def\be{\begin{eqnarray}}
\def\ee{\end{eqnarray}}

\def\bb{\bbox}

\begin{document}
\draft
 \title {\bf Two-neutron transfer in nuclei close to the dripline }

\author{E. Khan$^{a)}$,  N. Sandulescu$^{b),c)}$, Nguyen Van Giai$^{a)}$ and
M. Grasso$^{a)}$}

\vspace {03mm}

\address{
{\it a) Institut de Physique Nucl\'eaire, IN$_{2}$P$_{3}$-CNRS, 91406 Orsay,
France}\\
{\it b) Institute for Physics and Nuclear Engineering, P.O. Box MG-6, 76900
Bucharest, Romania}\\
{\it c) Royal Institute of Technology, SCFAB, SE-10691, Stockholm, Sweden}
}


\maketitle

\begin{abstract}

We investigate the two-neutron transfer modes induced by (t,p) reactions in
neutron-rich oxygen isotopes. The nuclear response to the pair transfer is
calculated in the framework of continuum-Quasiparticle Random Phase
Approximation (cQRPA). The cQRPA allows a consistent determination of the
residual interaction and an exact treatment of the continuum coupling.  The
(t,p) cross sections are calculated within the DWBA approach and the form
factors are evaluated by different methods : macroscopically, following the
Bayman and Kallio method, and fully microscopically. The largest cross
section corresponds to a high-lying collective mode built entirely upon
continuum quasiparticle states.

\end{abstract}

\vskip 0.5cm


\section{Introduction}

Two-neutron transfer reactions such as (t,p) or (p,t) have been used for
many years in order to study the nuclear pairing correlations (for a recent
review see Ref. \cite{oe01} ). The corresponding pair transfer modes are 
usually described in terms of pairing vibrations or pairing rotations 
\cite{be66,br73}.
High energy collective pairing modes, called giant pairing vibrations (gpv),
were also predicted \cite{br77,he80}, but they have not been observed yet.

Recently there is a renewed interest for the study of two-neutron transfer 
reactions with weakly bound exotic nuclei. These reactions would provide 
valuable information about the pairing correlations in nuclei far from 
stability. The use of two-neutron transfer reactions with exotic nuclei 
can also increase the chance of exciting the gpv mode, as discussed 
recently in Ref.\cite{fo02}.

The two-particle transfer modes are commonly described by the
particle-particle (pp)-RPA \cite{bo74,ri69} in the case of closed shell
nuclei and by the QRPA \cite{br73,fo02} in open shell nuclei. Most of the
cross section calculations use the distorted wave Born approximation (DWBA).
The form factor is usually calculated by means of macroscopic models
\cite{da85,da87} or by using the Bayman and Kallio method \cite{ba67}.
Several aspects of the model are under discussion, especially for absolute
cross section calculations
\cite{ig91} : one-step or sequential two-step
process, triton wave function, zero-range or finite-range DWBA. The
so-called 0s
\cite{br71}
approximation is also generally used to calculate the cross section in the DWBA
framework. In the latter approximation, the QRPA solutions act as a
spectroscopic factor
\cite{br73},
therefore the microscopic information does not affect the shape of the form
factor.

The calculation of the two-particle transfer modes in nuclei far from stability
presents additional difficulties compared to the case of stable nuclei. One of such 
difficulties is related to the continuum coupling, which becomes important in nuclei
close to the driplines. Therefore in
nuclei close to the driplines the pair transfer, the ground state properties and
the continuum coupling should be calculated consistently.

 The aim of this paper is to present such a consistent description of the
 two-particle transfer in the framework of the continuum-QRPA (cQRPA) 
 approach recently developed in Ref.\cite{kh02}. In the cQRPA the continuum
 is treated exactly and the residual interaction is derived from 
 the same effective force used in Hartree-Fock-Bogoliubov (HFB) for calculating the ground state properties.
 In this way the fluctuations of the particle and pairing densities,
 which are coupled together in the cQRPA, are calculated on the same footing
 with the ground state densities.

 The paper is organized as follows. In Section II we present the cQRPA model
 and we show how the response function for the two-particle transfer is 
 calculated within this model. In Section III we discuss the response function
 for the particular case of two neutrons transferred to the neutron-rich oxygen
 isotopes. In Section IV we present the calculation of cross sections
 for the transfer reaction $^{22}$O(t,p).

\section{The continuum-QRPA and the two-particle response}

Due to the concept of quasiparticle (qp), the QRPA unifies on the same
ground the particle-hole(ph)-RPA and the pp-RPA with the inclusion of the pairing
effects. In the continuum-QRPA model, presented in detail in
Ref.\cite{kh02}, the response of the nuclear system to an external
perturbation is obtained from the time-dependent HFB equations (TDHFB)
\cite{rs80}:

\begin{equation}\label{eq:tdhfb}
i\hbar\frac{\partial{\cal R}}{\partial t}=[{\cal H}(t) +
{\cal F}(t),{\cal R}(t)],
\end{equation}
where ${\cal R}$ and ${\cal H}$ are the time-dependent generalized density and
the HFB Hamiltonian, respectively. ${\cal F}$ is the external oscillating field :
\begin{equation}\label{eq:pert}
{\cal F} = F e^{-i\omega t} + h.c. .
\end{equation}
In Eq. (\ref{eq:pert}) $F$ includes both particle-hole and two-particle transfer
operators :
\begin{equation}\label{eq:extpart}
F=\sum_{ij} F^{11}_{ij} c_{i}^{\dagger}c_{j}+\sum_{ij}
(F^{12}_{ij} c_{i}^{\dagger}c_{j}^{\dagger}+ F^{21}_{ij} c_{i}c_{j}),
\end{equation}
and $c_{i}^{\dagger}$, $c_{i}$ are the particle creation and annihilation
operators, respectively.

In the small amplitude limit the TDHFB equations become:
\begin{equation}\label{eq:lin}
        \hbar\omega{\cal R}'=[{\cal H}',{\cal R}^0] + [{\cal H}^0,{\cal
	        R}']+[F,{\cal R}^0],
\end{equation}
where the superscript \ ' stands for the corresponding perturbed quantity. 

The variation of the generalized density ${\cal R}$' is expressed in
term of 3 quantities, namely $\rho'$, $\kappa'$ and $\bar{\kappa}'$,
which are written as a column vector : 

\begin{equation}\label{eq:rhodef}
{\bb{\rho'}}       =\left(
	\begin{array}{c}
	\rho' \\
         \kappa' \\
	 \bar{\kappa}' \\
        \end{array}
	\right),
	\end{equation}

where $\rho'_{ij} = \left<0|c^{\dagger}_jc_i|'\right>$
is the variation of the particle density,
$\kappa'_{ij} =\left<0|c_jc_i|'\right>$ and $\bar{\kappa}'_{ij} =
\left<0|c^{\dagger}_jc^{\dagger}_i|'\right>$ are the
fluctuations of the pairing tensor associated to the pairing vibrations and
$\mid ' \rangle$ denotes the change of the ground state wavefunction $|0>$
due to the external field. In contrast with the RPA where one needs
to know only the change of the ph density ($\rho'$), the variation of the
three quantities (\ref{eq:rhodef}) have to be calculated in the QRPA. In the
three dimensional space introduced in Eq. (\ref{eq:rhodef}), the first
dimension represents the particle-hole (ph) subspace, the second the
particle-particle (pp) one, and the third the hole-hole (hh) one. The
response matrix has therefore 9 coupled elements in QRPA, compared to one in
the RPA formalism.

The variation of the HFB Hamiltonian is given by:

\begin{equation}\label{eq:hvar}
\bb{{\cal H}}'=	\bf{V}\bb{\rho}',
\end{equation}
where $\bf{V}$ is the matrix of the residual interaction 
expressed in terms of the second derivatives of the HFB energy 
functional, namely:
					   
\begin{equation}\label{eq:vres}
{\bf{V}}^{\alpha\beta}({\bf r}\sigma,{\bf r}'{\sigma}')=
\frac{\partial^2{\cal E}}{\partial{\bf{\rho}}_\beta({\bf r}'{\sigma}')
\partial{\bf{\rho}}_{\bar{\alpha}}({\bf r}\sigma)},~~~\alpha,\beta = 1,2,3.
\end{equation}
In the above equation the notation $\bar{\alpha}$ means that whenever $\alpha$ is 2 or 3
then $\bar{\alpha}$ is 3 or 2.

Introducing for the external field the three dimensional column vector:

\begin{equation}\label{eq:f}
\bf{{F}}=	\left(\begin{array}{c}
	 {F}^{11} \\
	 {F}^{12}\\
	 {F}^{21}   \\
	\end{array}
	\right),
\end{equation}
the density changes can be written in the standard form:

\begin{equation}\label{eq:g}
{\bb{\rho'}}=\bf{G}\bf{F}~\boldmath \everymath{ },
\end{equation}
where $\bf{G}$ is the QRPA Green's function obeying the Bethe-Salpeter equation:

\begin{equation}\label{eq:bs}
\bf{G}=\left(1-\bf{G}_0\bf{V}\right)^{-1}\bf{G}_0=\bf{G}_0+\bf{G}_0\bf{V}\bf{G}.
\end{equation}
The unperturbed Green's function $\bf{G}_0$ has the form:

\begin{equation}\label{eq:g0}
{\bf{G}_0}^{\alpha\beta}({\bf r}\sigma,{\bf r}'{\sigma}';\omega)=
\sum_{ij} \hspace{-0.5cm}\int \hspace{0.3cm} \frac{{\cal U}^{\alpha 1}_{ij}({\bf r}\sigma)
\bar{{\cal U}}^{*\beta 1}_{ij}({\bf r}'\sigma')}{\hbar\omega-(E_i+E_j)+i\eta}
-\frac{{\cal U}^{\alpha 2}_{ij}({\bf r}\sigma)
\bar{{\cal U}}^{*\beta 2}_{ij}({\bf r}'\sigma')}{\hbar\omega+(E_i+E_j)+i\eta},
\end{equation}
where $E_i$ are the qp energies and ${\cal U}_{ij}$ are 3 by 2
matrices expressed in term of the two components of the HFB wave functions \cite{kh02}. 
The $\sum$ \hspace{-0.45cm}$\int$ ~ symbol in Eq. (\ref{eq:g0}) indicates that the
summation is taken both over the discrete and the continuum qp states.

The QRPA Green's function can be used for calculating the strength function
associated with various external perturbations. For instance the transitions from the
ground state to the excited states induced by a particle-hole external field
can be described by the strength function:

\begin{equation}\label{eq:stren}
S(\omega)=-\frac{1}{\pi}Im \int  F^{11*}({\bf r}){\bf{
G}}^{11}({\bf r},{\bf r}';\omega) F^{11}({\bf r}')
d{\bf r}~d{\bf r}'
\end{equation}
where ${\bf{G}}^{11}$ is the (ph,ph) component of the QRPA Green's function.
Examples of such calculations can be found in Ref. \cite{kh02}.

The quantity of interest in this work is the strength function describing the two-particle transfer from
the ground state of a nucleus with A nucleons to the excited states of a 
nucleus with A+2 nucleons. This strength function is :

\begin{equation}\label{eq:stren2}
S(\omega)=-\frac{1}{\pi}Im \int  F^{12*}({\bf r}){\bf{
G}}^{22}({\bf r},{\bf r}';\omega) F^{12}({\bf r}')
d{\bf r}~d{\bf r}'
\end{equation}
where ${\bf{G}}^{22}$ denotes the (pp,pp) component of the 
Green's function.  

\section{Pair transfer in oxygen isotopes: strength functions}

In the cQRPA model presented above one should calculate in the first step
 the ground state of the system within the continuum-HFB (cHFB) approach
 \cite{gr01}. The cHFB equations are solved in coordinate
 space assuming spherical symmetry. In the cHFB calculations presented here
 the mean field quantities are evaluated using the Skyrme interaction SLy4
 \cite{ch98}, while for the pairing interaction we take a zero-range
 density-dependent force. The parameters of the pairing force used here for
 calculating the neutron-rich oxygen isotopes are taken the same as in Ref
 \cite{kh02}. These parameters are fixed for a model space determined by a qp 
 energy cut-off equal to 50 MeV and a maximum angular momentum j=9/2.
 The HF single-particle and HFB qp energies corresponding to the
 $sd$ shell and to the $1f_{7/2}$ state are listed in Table \ref{tab:hf}. In
 both HF and cHFB calculations the state $1f_{7/2}$ is a wide resonance for
 $^{18-22}$O nuclei, while the state $1d_{3/2}$ is a narrow resonance. 

 In the cQRPA calculations we include the full discrete and continuum qp
 spectrum up to 50 MeV. These states, which generate a two-quasiparticle
 spectrum with a maximum energy of 100 MeV, are used to construct the
 unperturbed Green's function ${\bf G_{0}}$. The residual interaction is
 derived from the two-body force used in cHFB according to Eq.(2.7). The
 contribution given by the velocity-dependent terms of the Skyrme force to
 the residual interaction is calculated in the Landau-Migdal approximation.
 Due to this approximation the self-consistency of the HFB+QRPA equations is
 not exactly preserved in the numerical calculations. Therefore, the
 spurious mode associated with the center-of-mass motion is not at zero
 energy.  In order to put the spurious state energy to zero we renormalize
 the residual interaction by a factor equal to 0.80 for all the calculated
 isotopes. The Goldstone mode is also located at zero energy in the ph
 monopole response with this renormalization.

 The strength function for the two-neutron transfer is calculated using Eq.
 (\ref{eq:stren2}). For the radial function $F^{22}(r)$ we take the form $r^L$,
 which is equal to the unity for the L=0 pair transfer mode considered here \cite{fo02}.
 The unperturbed Green's function is calculated with an averaging interval
 $\eta$ equal to 0.15 (1.0) MeV for excitations energies below (above) 11
 MeV.

 The results for the strength function corresponding to a neutron pair
transferred to the oxygen isotopes $^{18,20,22}$O are shown in Fig.
\ref{fig:oxystrength}. The exact continuum treatment is also compared
to box discretization calculations (the box radius is 22.5 fm, and the
averaging interval $\eta$ is 0.15 MeV). For the isotopes $^{18,20}$O the
first three peaks correspond to a pair transferred mainly to the states
$1d_{5/2}$, $2s_{1/2}$ and $2d_{3/2}$, respectively. For the isotope
$^{22}$O the subshell $d_{5/2}$ is essentially blocked for the pair
transfer. Therefore in this nucleus we can clearly identify only two peaks
below 11 MeV, corresponding to a pair transferred to the states $2s_{1/2}$
and $2d_{3/2}$. For all the isotopes we can see a broad resonant structure
around 20 MeV which is built mainly upon the single-particle resonant sate
$1f_{7/2}$. This two-quasiparticle broad resonance has the characteristics
of a giant pairing vibration \cite{br77,he80,fo02}. It should be noted that
the continuum treatment affects the magnitude of the lowest state for the
three responses. This is due to the collective nature of this state, since
unbound configurations such as the (1d$_{3/2}$)$^2$ contribute
to this low-lying state. This points to the necessity to use exact
continuum calculations even to predict transitions towards low-lying states.
The state at 9.8 MeV on the $^{22}$O + 2n spectrum is embedded in the
continuum and it is naturally more affected by the continuum treatment.

 The influence of the residual interaction on the pair transfer modes is
illustrated in Fig. \ref{fig:o22trength} for the case of $^{22}$O+2n,
emphasizing the collective nature of the pairing response and the difference
between pairing rotations and pairing vibrations
\cite{br73}. As expected, the residual interaction shifts down the position of the
two-quasiparticle resonant states and increases their strengths. Pairing
rotations occur for open shell nuclei when many 0$^+$ pairs of nucleon are
feeding the ground state, forming a superfluid condensate. This should be the case
for the G$_0$ response where a pair is added in a pure two-quasiparticle
configuration, increasing the number of pairs of the superfluid condensate.
Pairing vibrations are collective states where the pair contributions are
mixed by the residual interaction and they should therefore be described by the G
response. In the $^{22}$O + 2n case, pairing vibrations are much stronger
than pairing rotations, due to the relatively small number of pairs
(\verb1~14) which contribute to the superfluid condensate to form a pairing
rotation mode. The first two peaks located at 2.1 MeV and 10.8 MeV
in the G$_0$ response correspond to the addition of two
neutron qp on the (2s$_{1/2}$)$^2$ and (1d$_{3/2}$)$^2$ subshells,
respectively. Apart from that, we can also notice a sensitive change in the
spreading widths of the two-quasiparticle resonant states when the residual
interaction is turned on. Thus, due to the mixing of the configurations
$(1f_{7/2})^2$ and $(1d_{3/2})^2$ by the residual interaction, the broad
peak around 18 MeV becomes narrower and the narrow peak around 10 MeV
becomes wider. This is a general effect which appears whenever in the
two-body wave functions wide and narrow single-particle resonant states are
mixed together \cite{bet02} .

\section{Pair transfer in oxygen isotopes: cross sections}

The DWBA calculation of the cross section for the two-neutron transfer
requires the form factor, which represents the correlation between
the two neutrons and the initial nucleus \cite{sa83}.
In order to compare the influence of various approximations, the form 
factor is calculated below by three different ways : macroscopically, by 
the so-called
Bayman and Kallio method, and using the cQRPA model. The calculations based on
the cQRPA allows for a straightforward study of the effects of the pairing
correlations and of the continuum coupling upon the cross section. The form factors 
and the cross sections will be calculated below for the particular transfer 
reaction  $^{22}$O(t,p).

\subsection{The macroscopic form factor}

In Refs.\cite{da85,da87} the form factor is calculated from the variation
of the optical potential U with respect to the change of the number of
particles:

\begin{equation}\label{eq:fmacr}
F(r)= \beta_p \frac{dU}{dA}
\end{equation}
where $\beta_p$ is the so-called pairing deformation parameter representing the
strength of the two-neutron pairing transfer reaction.
Using for the nuclear radius the relation $R=r_0 A^{1/3}$ one gets :

\begin{equation}\label{eq:fmacr2}
F(r)= \beta R \frac{dU}{dr}
\end{equation}
where

\begin{equation}\label{eq:beta}
\beta=\frac{\beta_p}{3A} 
\end{equation}
Eq. (\ref{eq:fmacr2}) is referred to the macroscopic form factor.
In this model for the form factor the transfer is considered as an inelastic
process corresponding to a deformation parameter given by Eq.
(\ref{eq:beta}). One advantage of this model is that the two-particle transfer
cross section can be calculated by knowing only the optical potential in the
entrance channel. It should be stressed that Eq. (\ref{eq:fmacr}) assumes that 
the optical potential changes smoothly with the number of nucleons. 
Therefore, one expects that this model for the form factor works better 
for heavy nuclei.

The cross section for the reaction $^{22}$O(t,p) is calculated in the DWBA
approximation using the ECIS88 code \cite{ra81}. The reaction energy is 
chosen to be 15 MeV/A. For the optical potential corresponding to
the  system $^{22}$O+t we use the potential derived by Becchetti and 
Greenlees \cite{be71}. The parameters corresponding to this potential are
summarized in  Table \ref{tab:bg}. The $\beta_p$ parameter associated to
a particular transfer mode is obtained by taking the average of the square
root of the integral of the strength function over the energy region 
corresponding to that mode. The calculations are performed for the first two peaks
located at 1.6 MeV and 9.8 MeV and for the broad resonant region located
around 18 MeV. The angular distribution corresponding to the state at 1.6 MeV,
displayed in  Figure \ref{fig:macrodis}, is showing a typical diffraction pattern.
For the other states the pattern of the angular distribution is the same. 
The only difference is in their magnitude, which depends on the 
$\beta_p$ value.  Table \ref{tab:sigmamacro} shows the total  cross-sections obtained for 
the three states mentioned above. The continuum treatment affects
all the cross-section by 20\%. One notices that the cross section for the
 broad resonant structure located around 18 MeV is about  5 times 
larger than the cross section corresponding to the state at E=1.6 MeV.

\subsection{The Bayman and Kallio form factor}

Bayman and Kallio have proposed a method to calculate the form factor on a
semi-microscopic ground. In this method the two-particle wave function of the
transferred pair is expressed in term of the single-particle wave functions
corresponding to a Woods-Saxon potential \cite{ba67}. The spectroscopic 
factors of the single-particle states are an input of the calculation.

An example of such calculation for the form factor is shown in Fig.
\ref{fig:formfactbk}. The calculation is for the configuration
$(2s_{1/2})^2$ and for the $^{22}$O+2n system. The Woods-Saxon potential was
obtained by the separation energy method, where the depth of the potential
is set in order to reproduce the binding energy of the single-particle to
the core. The radius and the diffuseness of the potential were taken at the
standard values of 1.25 fm and 0.65 fm, respectively. For the spectroscopic
factor of the single-particle state we take the value 1. As it can be seen
in Fig. \ref{fig:formfactbk}, the Bayman and Kallio form factor has also
negative values. This is due to the fact that in this calculation the two
neutrons are not considered at the same position in the two-neutron wave
function.


The previous $^{22}$O+t Becchetti and Greenlees optical potential
\cite{be71} is used for the entrance channel and the $^{22}$O+p Becchetti
and Greenlees \cite{be69} for the exit channel, in order to calculate the
DWBA cross-section. The resulting optical potential parameters are given in
table \ref{tab:bg}. The DWBA calculations are performed with the DWUCK4
\cite{ku} code and using the zero-range approximation. In this approximation the
two-neutrons and the residual fragment are located at the same point and
the range function
is expressed through a simple constant D$_0$ \cite{sa83}.
 For the (t,p) reaction we take D$_0$ = 2.43 10$^4$ MeV$^2$ fm$^3$ \cite{van}. 
 This value relies on measurements of the 2n+p system and may be subject to
 uncertainties \cite{sa83}.
 The angular distribution for the (t,p) 
 reaction at E=15MeV/A obtained with the Bayman and Kallio form factor
 is shown in Fig. \ref{fig:microdis}. 
 As discussed in Refs. \cite{br73,ig91,sa83}, the shape of the angular 
distribution
 is usually described correctly by the zero-range approximation, but not its
 magnitude, which is generally underestimated by a large amount. Therefore in
 what follows we will focus our discussion not on the absolute values of
 the cross sections, but rather on the relative values obtained using
 different form factors.

\subsection{The microscopic form factor}

In the nuclear response theory the transition from the ground state to 
the excited  state $|\nu\rangle$  of the same nucleus is determined by
the transition density defined by:

	\begin{equation}\label{eq:rhor}
\rho^\nu\left({\bf r}\sigma\right) = 
	\left<0|c^{\dagger}\left({\bf r}\sigma\right)
	c\left({\bf r}\sigma\right)|\nu\right>
\end{equation}
where $c^{\dagger}\left({\bf r}\sigma\right)$ is
the particle creation operator in coordinate space.

The corresponding quantity for describing pair transfer processes is
the pair transition density defined by:

\begin{equation}\label{eq:rhotildr}
        \kappa^\nu\left({\bf r}\sigma\right) = 
	\left<0|c\left({\bf r}\bar{\sigma}\right)
	c\left({\bf r}\sigma\right)|\nu\right>
\end{equation}
where $c^{\dagger}\left({\bf r}\bar{\sigma}\right)$=
$-2\sigma c^{\dagger}\left({\bf r}-\sigma\right)$ is its time reversed
counterpart. The pair transition density defined above determines the transition
from the ground state of a nuclei with A nucleons to a state $|\nu\rangle$ of a 
nucleus with A+2 nucleons. This quantity is the output of cQRPA calculations.

The form factor for the pair transfer is obtained by folding the pair
transition density $\kappa^\nu$ (Eq. (\ref{eq:rhotildr})) with the interaction 
acting between the transferred pair and the residual fragment \cite{oe01}.
In the zero-range approximation used here the dependence of this interaction
on the relative distance between the pair and the fragment is taken as a delta
force. Therefore in this approximation the pair transition density 
(\ref{eq:rhotildr}) coincides with the form factor \cite{sa83}.

Fig. \ref{fig:formfactbk} displays the pair transition density corresponding
 to the state around 2 MeV in the $^{22}$O+2n system. The dotted line
 represents the QRPA results while the dashed line shows the results derived
 from the {\bf G$_0$} response, i.e.  without taking into account the
 residual interaction between the quasiparticles. Thus the dashed curve
 corresponds to a pair transferred in the pure two-quasiparticle
 configuration (2s$_{1/2}$)$^2$. On the other hand the Bayman and Kallio
 form factor, shown by the full line, corresponds to the addition of a pair
 of neutrons in a pure two-particle configuration (2s$_{1/2}$)$^2$. As can
 be seen in Fig. \ref{fig:formfactbk}, the form factor derived from the {\bf
 G$_0$} response has at large distances a smaller amplitude compared to the
 Bayman and Kallio form factor. This is due to the occupancy of the state
 2s$_{1/2}$, which is different from zero in the case of the {\bf G$_0$}
 response. Another reason is that in the pair transition density given by
 (\ref{eq:rhotildr}) the two neutrons are taken to be at the same position
 whereas in the Bayman and Kallio method the two neutrons are allowed to be
 at different positions. From Fig. \ref{fig:formfactbk} one can also see that
 the effect of the residual interaction is to increase the magnitude of the
 form factor, which is consistent with the increase of the strength shown in
 Fig. 1.

Fig. \ref{fig:microdis} displays the angular distributions for
the reaction  $^{22}$O(t,p)  at 15 MeV/A. The optical potential
we have used in calculations is the same as above. The angular
distributions are mainly sensitive to the surface part of the form factor,
where the form factors calculated with the Green's functions {\bf G} and 
{\bf G$_0$} show strong variations. The diffraction minima are shifted by 5 to 10 degrees
 between the two microscopic calculations. The angular distribution calculated
 with the Bayman and Kallio form factor drops faster with increasing angle. 
This is due to the spatial extension of the two transferred neutrons which 
produces the negative part of the form factor.

In order to see the continuum effect on the form factor, in Fig.
\ref{fig:fomrfactconti} we
display the transition densities calculated by solving the QRPA equations
with continuum and box-type asymptotic conditions.
The corresponding effect of the continuum treatment on the (t,p) angular distribution 
is shown in Fig. \ref{fig:microdisconti}. As expected, the effect
is small for the transition towards the state
located at 1.6 MeV, built mainly upon bound qp states, 
and large for the state located at 9.8 MeV, which is built mainly 
upon narrow resonant qp states. From Fig. \ref{fig:fomrfactconti}
we can see that the continuum treatment has also some effect on the form
factor corresponding to the high energy mode around 18 MeV, especially at
small values of the nuclear radius. However its  effect on the global cross
 section remains negligible, about 3\%.

\section{Conclusions}

In this paper we have investigated the pair transfer in neutron-rich oxygen
isotopes in the framework of the continuum-QRPA. The form factors are
calculated with a macroscopic model, the Bayman and Kallio approach, and
fully microscopically. The cross section is evaluated by using the DWBA and
the zero-range approximation. The response function exhibits some narrow
resonances corresponding to a pair transfer in the single-particle states of
$sd$ shell and a broad peak at high energies. This peak is built mainly upon
the single-particle resonance $1f_{7/2}$ and its cross section is much
larger than the one associated to the lower energy transfer modes. Since
this high energy transfer mode is formed mainly by single-particle states
above the valence shell, this mode is similar to the giant pairing
vibration mode suggested long ago. Although such a mode has not been
detected yet, the pair transfer reactions involving exotic loosely bound
nuclei may offer a better chance for this undertaking. On the theoretical
side, one needs to make a better estimation for the absolute cross section
associated to this mode. Due to the collectivity of the final states the
continuum treatment has also an impact on low-lying states. The form factors
calculated microscopically show an effect of the continuum
treatment on the form factor for high energy states. The angular
distributions are mainly affected in their diffraction minima for narrow
high energy states. In the case of the gpv, no drastic influence of the
exact continuum treatment is observed. It should be interesting to perform
the DWBA calculations without the zero-range approximation in order to make
more quantitative predictions for the cross section. This requires to fold
the pair transition densities with the interaction between the two neutrons
and the residual fragment.

\vskip 0.5cm

\noindent{\bf Acknowledgments} E. K. acknowledges fruitful discussions with
S. Fortier and J. Van de Wiele. N.S. thanks Roberto Liotta for
useful discussions on giant pairing vibrations and to the Swedish 
Foundation for International Cooperation in Research and Higher Education 
(STINT) for financial support.

\newpage

\begin{table}[h]
\begin{tabular}{c|c||c|c|c|}
   &     & $^{18}$O   & $^{20}$O  & $^{22}$O
\\ \hline
1d$_{5/2}$ & HF & -6.7 & -7.0 & -7.45
\\ \hline
1d$_{5/2}$ & cHFB & 2.26 & 2.08 & 2.30
\\ \hline
\\ \hline
2s$_{1/2}$ & HF & -4.0 & -4.2 & -4.6
\\ \hline
2s$_{1/2}$ & cHFB & 3.46 & 2.28 & 1.05
\\ \hline
\\ \hline
1d$_{3/2}$ & HF & (0.46;0.02) & (0.51;0.03) & (0.42;0.02)
\\ \hline
1d$_{3/2}$ & cHFB & (7.74;0.12)  & (6.60;0.29) & (5.39;0.01)
\\ \hline
\\ \hline
1f$_{7/2}$ & HF & (5.50;1.35) & (5.24;1.24)  &  (4.86;1.04)
\\ \hline
1f$_{7/2}$ & cHFB &(12.77;1.13)  & (12.14;0.83) & (10.05;0.69)
\\ \hline
\\ \hline
\end{tabular}

\caption{\label{tab:hf}}
$^{18,20,22}$O single-particle and single quasi-particle neutron energies calculated
in the Hartree-Fock and the continuum-HFB models, respectively. The Skyrme
interaction used is SLy4 and the resonances are displayed in brackets :
(energy;width).
\end{table}

\begin{table}[h]
\begin{tabular}{c||c|c|c||c|c|c||c|c|c||c|c|c|}
 &  V   &  r$_0$ & a$_0$ &
  W   &  r$_W$ & a$_W$ &
  W$_S$   &  r$_{W_S}$ & a$_{W_S}$ &
  V$_{so}$   &  r$_{so}$ & a$_{so}$ 
\\ \hline
$^{22}$O+t  & 156 & 1.20 & 0.72 & 28 & 1.40 & 0.84 & 0 & 0 & 0 & 2.5 & 1.20 & 0.72
\\ \hline
$^{24}$O+p  &50.6 & 1.17 & 0.75 & 5.9 & 1.32 & 0.74 & 6.0 & 1.32 & 0.74 & 6.2
& 1.01 & 0.75
\end{tabular}

\caption{\label{tab:bg}}
The parameters of the optical potential \cite{be71,be69} used for the transfer
reaction $^{22}$O(t,p) at 15 MeV/A. V is the depth used for the real part of the potential,
W for the imaginary absorptive part, W$_s$ for the surface imaginary part and V$_{so}$ for
 the spin-orbit part. The depths are given in MeV and the radii and diffuseness parameters in fm.
\end{table}

\begin{table}[h]
\begin{tabular}{|c|c|c|c|}
 &E=1.6 MeV & E=9.8 MeV & gpv
\\ \hline
Box & 0.062 & 0.147 & 0.316
\\ \hline
Cont & 0.052 & 0.122 & 0.270
\end{tabular}

\caption{\label{tab:sigmamacro}}
Total cross section (mb) of the $^{22}$O(t,p) reaction at E=15 MeV/A.
The calculations are performed for the first two resonant states of the $^{22}$O+2n
system and for the giant pairing vibration (gpv) mode. The results are 
obtained using the macroscopic model for the form factor \cite{oe01}.
The first (second) row shows the results obtained with box (continuum)
boundary conditions.

\end{table}

\begin{figure}[h]
\vspace{0.0cm}
\caption {The QRPA response for the two-neutron transfer on
 $^{18,20,22}$O. The exact continuum calculations are in solid lines
whereas the calculations with box boundary conditions are in
dashed lines. The results are displayed as functions of E$^*$, the excitation energy with respect to the parent nucleus ground state.} 
\label{fig:oxystrength}
\end{figure}

\begin{figure}[h]
\vspace{0.0cm}
\caption {The response function for the two-neutron transfer on $^{22}$O. 
The unperturbed response is in solid line and the QRPA response
in dashed line.}
\label{fig:o22trength}
\end{figure}

\begin{figure}[h]
\vspace{0.0cm}
\caption { DWBA calculations for the reaction $^{22}$O(t,p) at 15 MeV/A
calculated with a macroscopic form factor (see text for details).}
\label{fig:macrodis}
\end{figure}

\begin{figure}[h]
\vspace{0.0cm}
\caption {The microscopic form factors for the state located close to 2 MeV in the
 $^{22}$O+2n system. The solid line is obtained using the Bayman and Kallio method,
 the dashed line with the unperturbed {\bf G$_0$} response, and the dotted line with 
 the QRPA response.}
\label{fig:formfactbk}
\end{figure}

\begin{figure}[h]
\vspace{0.0cm}
\caption {15 MeV/A $^{22}$O(t,p) DWBA calculations using the form factors shown in Fig. 4.}
\label{fig:microdis}
\end{figure}

\begin{figure}[h]
\vspace{0.0cm}
\caption {The QRPA form factors for the states located at 1.6 MeV, 9.8 MeV 
and at the giant pairing vibration region. The form factor obtained with
box boundary conditions are in solid lines whereas the form factors obtained
with the exact continuum treatment  are displayed in dashed lines. The calculations
are done for the  $^{22}$O+2n system. }
\label{fig:fomrfactconti}
\end{figure}

\begin{figure}[h]
\vspace{0.0cm}
\caption {DWBA calculations for the states located at 1.6 MeV (up) 
 and 9.8 MeV (down). The solid (dashed) line corresponds to the QRPA results
 obtained with box (exact) boundary conditions. The calculations corresponds
 to the system  $^{22}$O+2n. }
\label{fig:microdisconti}
\end{figure}



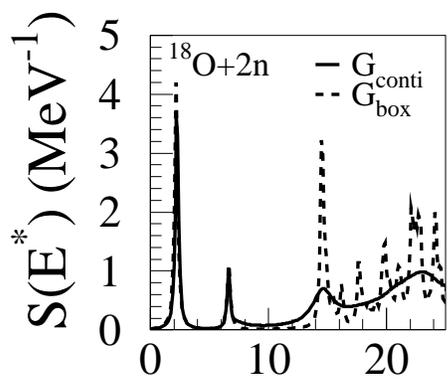
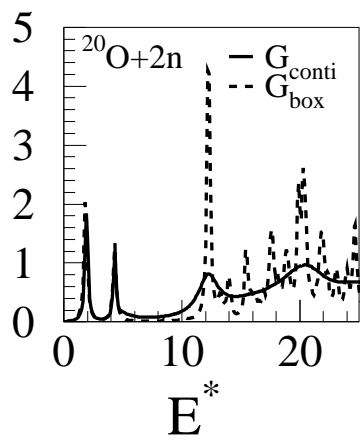
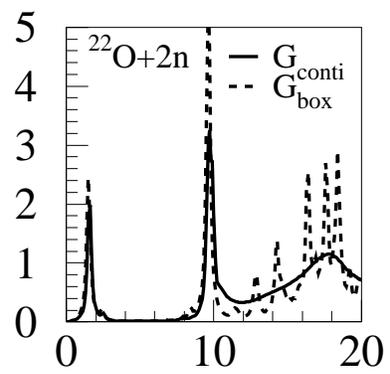
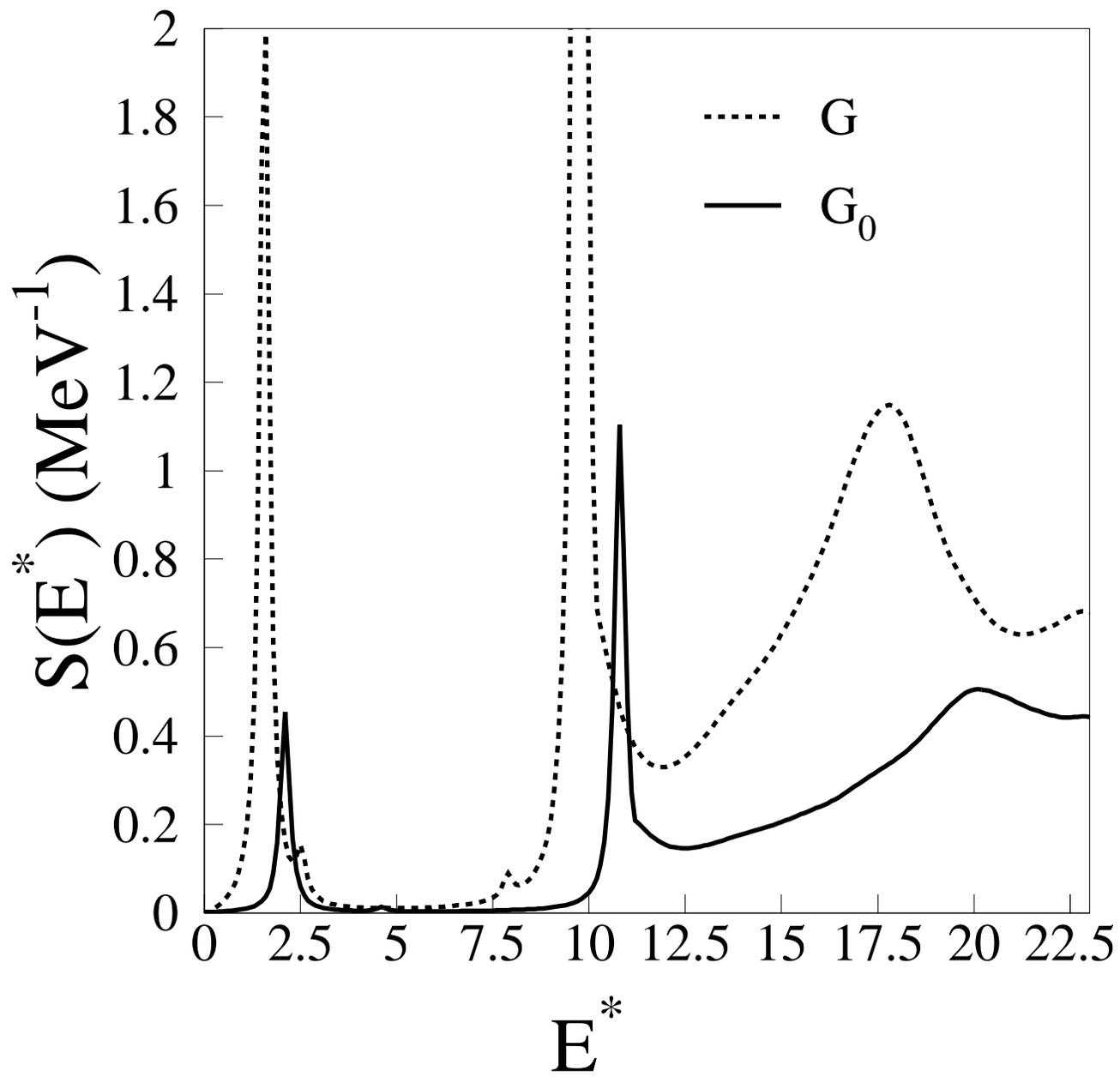
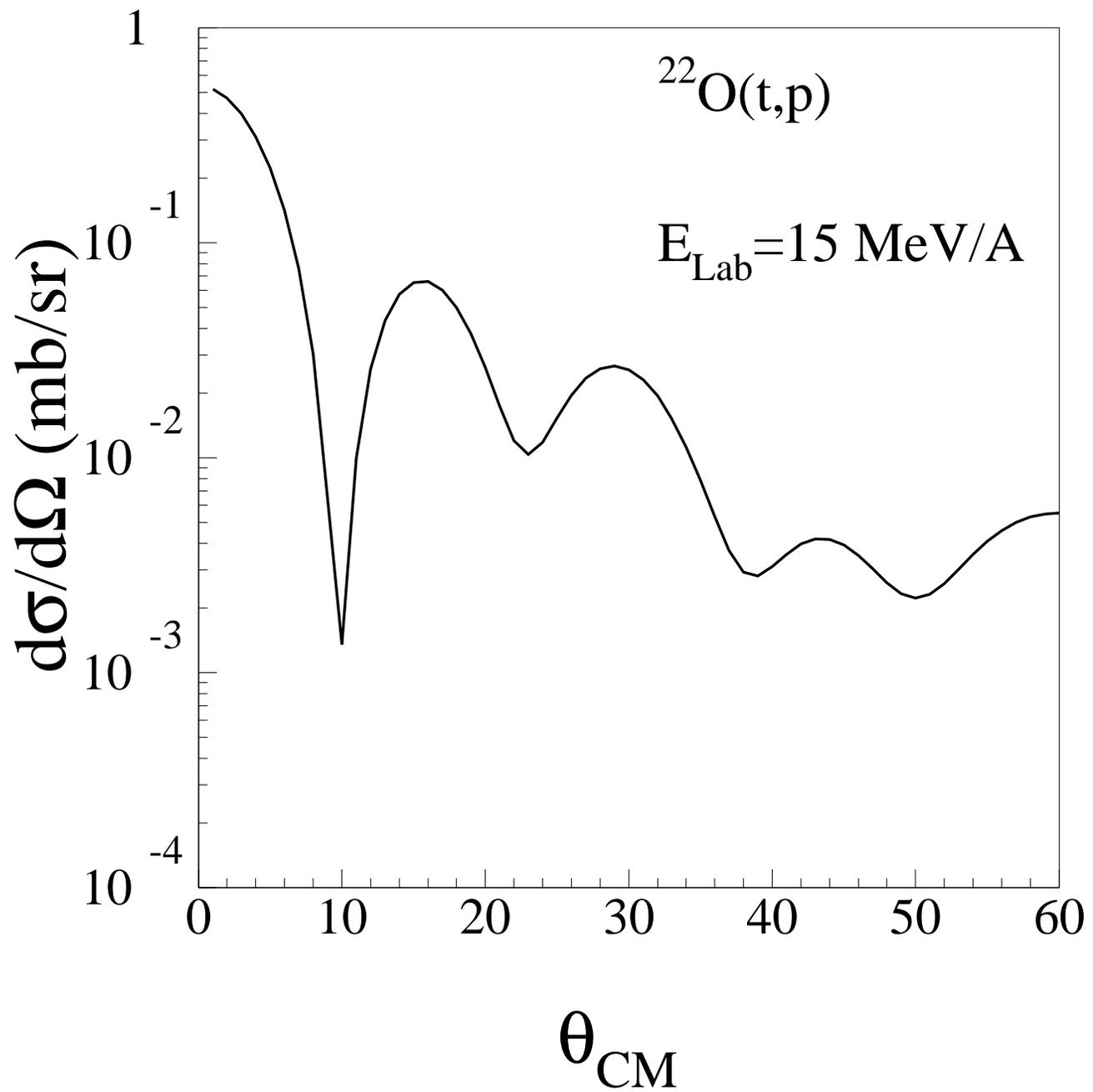
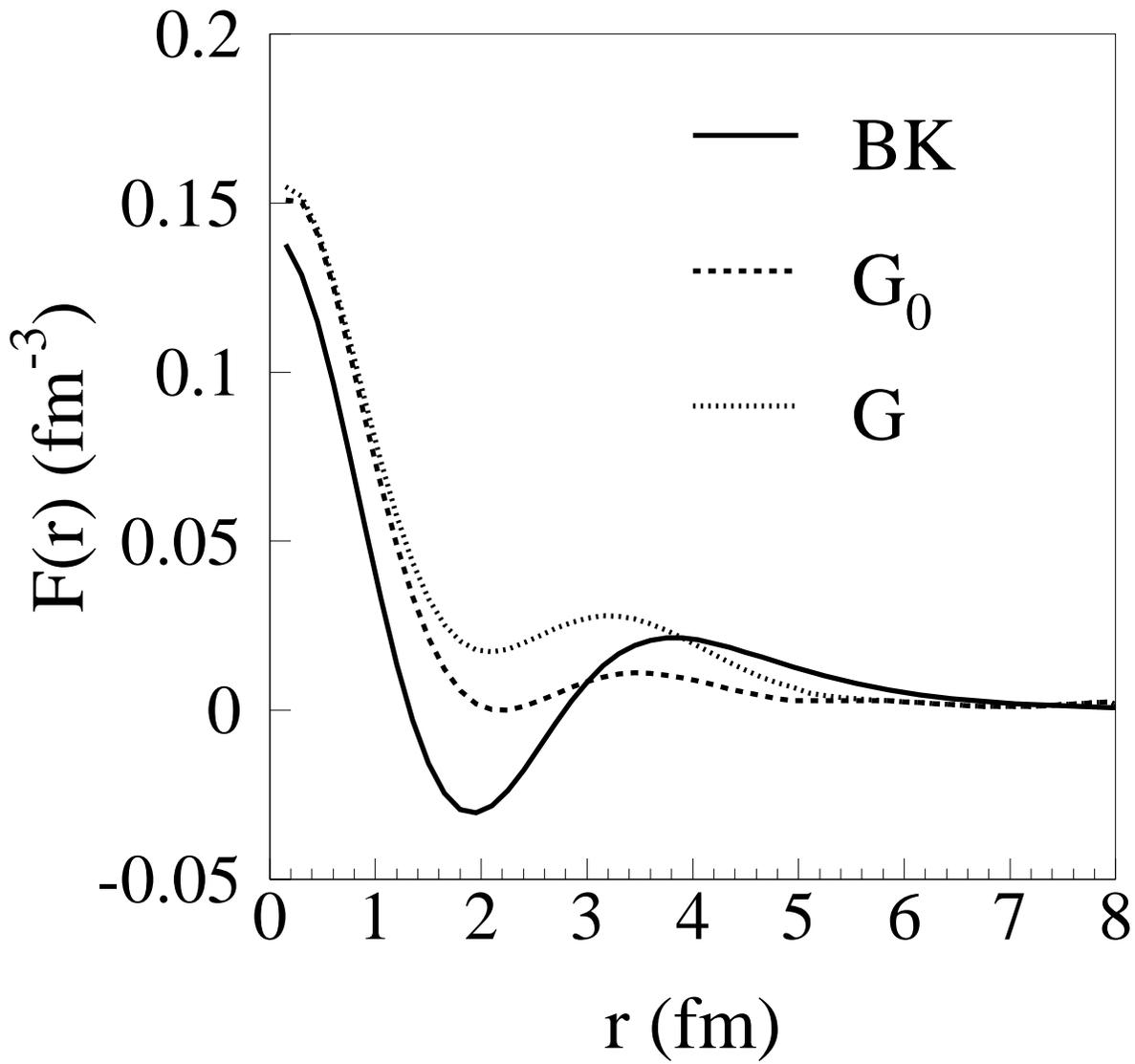
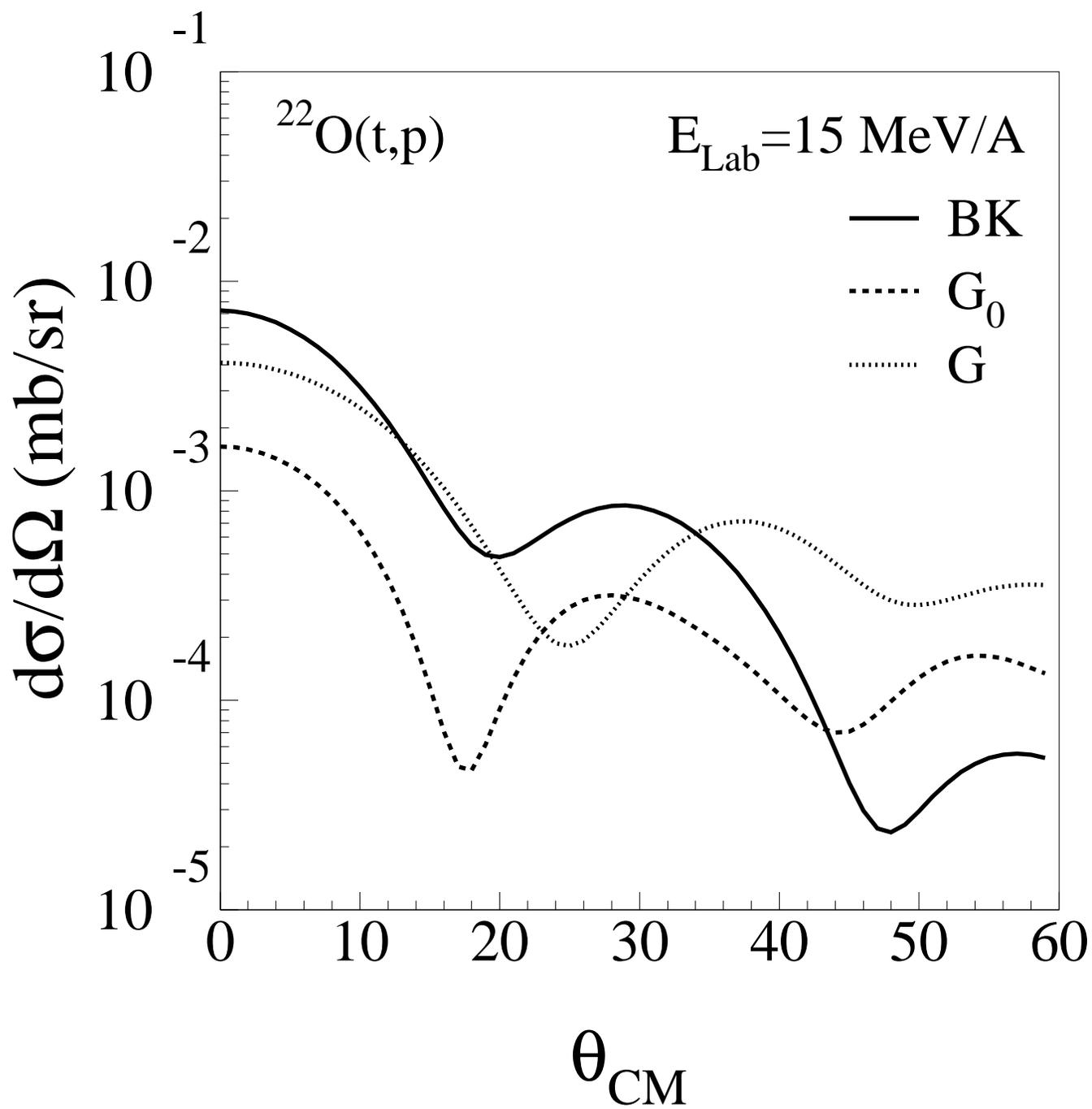
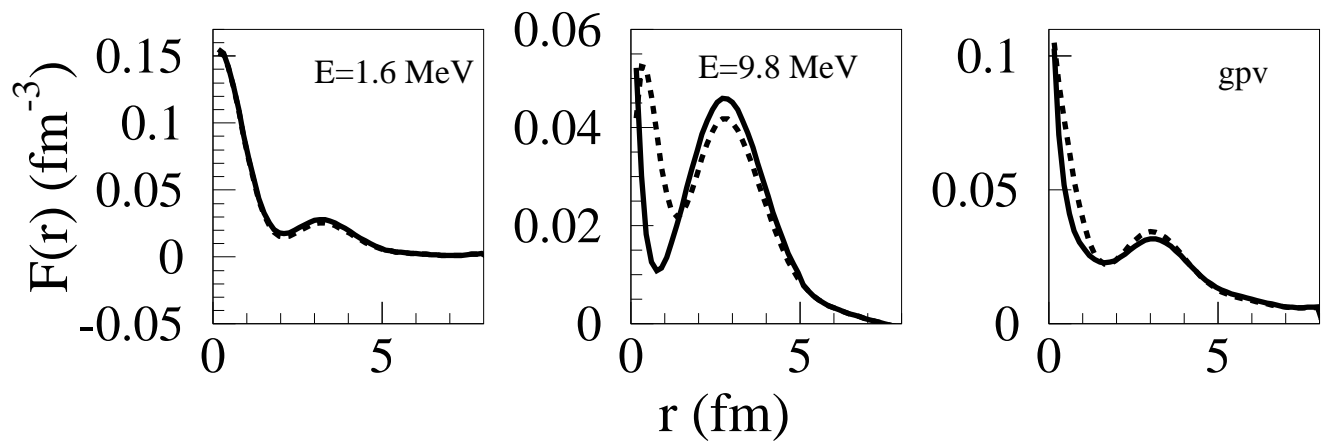
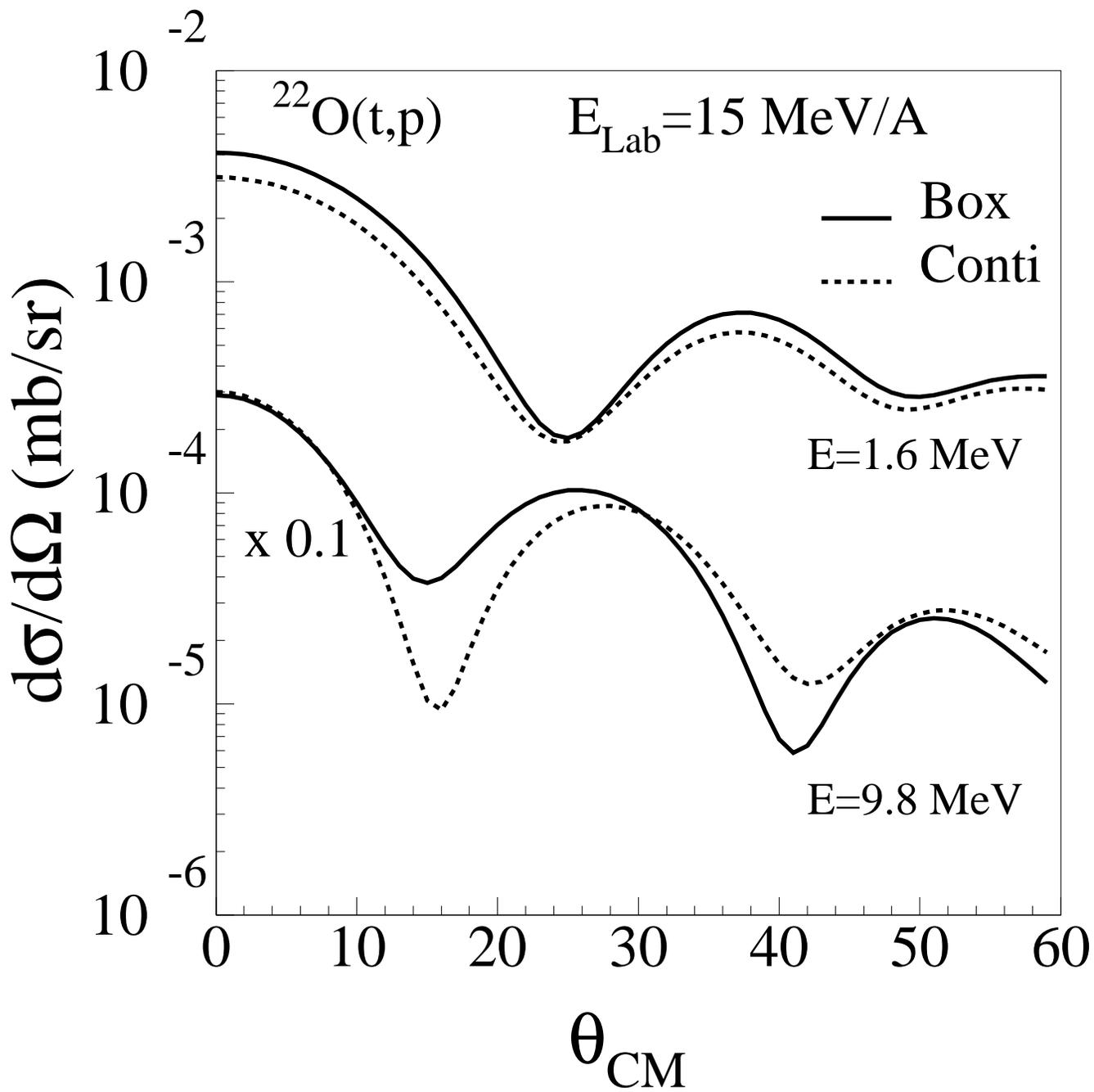
\end{document}